\documentclass[12pt, a4]{article}
\usepackage{txfonts}
\usepackage{bm}
\usepackage[dvipdfmx]{graphicx}
\usepackage{amstext}
\usepackage{here}
\usepackage{authblk}

\title{Magnetism and high--magnetic field magnetization in alkali superoxide CsO$_2$
}

\author{
Mizuki Miyajima$^{1}$\thanks{p2qm1zzv@s.okayama-u.ac.jp}, 
Fahmi Astuti$^{2,3}$, 
Takeshi Kakuto$^1$, 
Akira Matsuo$^4$, 
Dita P. Sari$^{2,5}$, 
Retno Asih$^{2,5}$, 
Kouichi Okunishi$^6$, 
Takehito Nakano$^5$, 
Yasuo Nozue$^5$, 
Koichi Kindo$^4$, \\
Isao Watanabe$^{2,3,5}$ and 
Takashi Kambe$^{1}$\thanks{kambe@science.okayama-u.ac.jp}
}

\affil{$^1$Department of Physics, Okayama University, Okayama 700-8530, Japan \\
$^2$ Advanced Meson Science Laboratory, RIKEN Nishina Center, Wako, Saitama 351--0198, Japan \\
$^3$ Department of Physics, Hokkaido University, Sapporo 060--0808, Japan \\
$^4$ Institute for Solid State Physics, The University of Tokyo, Kashiwa, Chiba 277--8581, Japan \\
$^5$ Department of Physics, Osaka University, Toyonaka, Osaka 560--0043, Japan \\
$^6$ Department of Physics, Niigata University, Niigata 950--2181, Japan \\
} %\\

\date{}

\begin{document}
\maketitle
\newpage
\begin{abstract}
Alkali superoxide CsO$_2$ is one of candidates for the spin--$\frac{1}{2}$ one--dimensional (1D) antiferromagnet, which may be sequentially caused by an ordering of 
the $\pi$--orbital of O$_{2}^-$ molecule below $T_{\rm S} \sim 70$ K. 
Here, we report on the magnetism in powder CsO$_2$ and high--magnetic field magnetization measurements 
in pulsed-magnetic fields of up to 60 T. 
We obtained the low temperature phase diagram around the antiferromagnetic ordering temperature $T_{\rm N} = 9.6$ K 
under the magnetic field. 
At 1.3 K, remarkable up--turn curvature in the magnetization around a saturation field of $\sim 60$ T is found, indicating 
the low--dimensional nature of the spin system. 
The saturated magnetization is also estimated to be $\sim 1 \mu_{\rm B}$, which corresponds to the spin--$\frac{1}{2}$. 
We will compare it with the theoretical calculation. \\
\end{abstract}

%\section{Introduction}
O$_2$ is a magnetic molecule with a spin--1,  originating with two unpaired electrons on antibonding $\pi^*$ orbitals. 
The magnetism of three structural phases of solid O$_2$ has been studied long ago \cite{Hemert, Uyeda}. 
The magnetic exchange interaction between O$_2$ molecules depends strongly on their distance as well as relative displacement 
\cite{Hemert, Uyeda}. 
O$_2$ molecules adsorbed in a porous metal complexes have been investigated by a magnetic susceptibility 
and high magnetic field magnetization \cite{kobayashi, hori}. 
The O$_2$--array is confined in the nano--channel region and exhibits a meta--magnetic transition under high--magnetic field, 
which is considered to be due to a configurational transition of O$_2$ dimer. 
Other candidate for O$_2$ molecular based magnet  is achieved by a encapsulation of O$_2$ molecules 
in single--walled carbon nanotubes. 
From the magnetic susceptibility and high--magnetic field magnetization, it is proposed as a spin--1 Haldane state \cite{hagiwara}. 
Since then, the magnetism of O$_2$ molecule has been attracted much attention. 

Magnetism of alkali superoxide, AO$_2$ (A = Na, K, Rb and Cs), originates in unpaired $\pi$--electron on the O$_2^-$ molecular anion, 
where one electron transfers from the alkali metal to O$_2$ making a spin--$\frac{1}{2}$ state. 
The transferred electron has a freedom which orbital ($\pi^*_x$ or $\pi^*_y$) to select and the orbital ordering would be expected to realize 
the three--dimensional magnetic exchange pathways. 
At room temperature, it is proposed that KO$_2$, RbO$_2$ and CsO$_2$ have the same tetragonal crystal structure  (I4/mmm) while 
NaO$_2$ has a cubic crystal structure (Fm$\bar3$m). 
Recently, CsO$_2$ has been attracted considerable attention because of the one--dimensional (1D) antiferromagnetic (AF) nature \cite{Blake}. 
Below the structural phase transition temperature $T_{\rm S} = 70$ K, the magnetic susceptibility follows a well--known Bonner--Fisher curve. 
It is proposed that the structural phase transition is accompanied with the $\pi$--orbital ordering  of O$_2$ molecules, which leads to a zig--zag like 1D chain along the $b$--axis. 
If this is the case, the 1D AF super--exchange pathway might be via the Cs atom. 
NMR experiment shown a power--law decay in inverse spin--lattice relaxation rate at low temperature, 
suggesting that the ground state of CsO$_2$ is a Tomonaga--Luttinger liquid (TLL) state \cite{Klanjsek}. 
At lower temperatures, it was proposed that CsO$_2$ showed an AF transition at $T_{{\rm N}} = 9.6$ K \cite{Blake, Hesse}, but 
the low temperature magnetism of CsO$_2$ have not yet been clarified. 

KO$_2$ and RbO$_2$ also showed AF transitions at $T_{\rm{N}} = 7$ and 15 K, respectively \cite{Hesse}. 
In spite of the same crystal structures as CsO$_2$ at room temperature, 
their magnetic susceptibilities show Curie--Weiss behavior from room temperature to $T_{\rm{N}}$, but do not show the 1D behaviors. 
As the low--temperature structures  in KO$_2$ and RbO$_2$ have not been settled, the magnetic exchange pathway have not been determined yet. 
Moreover, in NaO$_2$, both the low temperature structure and the magnetic ground state is still under debate. 
Therefore, clarifying how the structure changes at low temperature is important subject to 
understand the correlation between the orbital ordering and the magnetism. 

Accordingly, alkali superoxide is one of the fascinating candidates for molecular based low--dimensional magnet 
and may have a strong coupling between spin and orbital degrees of freedom. 
In this Letter, we have shown the low--temperature magnetism of CsO$_2$, especially, high--magnetic field magnetization for the first time. 
We will present the full temperature dependence of magnetization around $T_{\rm N}$ in order to discuss the magnetic phase diagram. 
To discuss the dimensionality of CsO$_2$, the high--field magnetization will be compared with the theoretical calculation. 

%\section{Experiments}
We synthesized CsO$_2$ powder using a liquid ammonia method. 
Alkali--metal was placed in a glass tube in a Ar--filled glove box (O$_2$ and H$_2$O $< 0.1$ ppm) that was then dynamically pumped down to 10$^{-2}$ Pa. 
The glass tube was cooled by liquid N$_2$ in order to condense the NH$_3$. 
After the glass tube was filled with liquid NH$_3$ (typically, $\sim 10$ ml), O$_2$ gas was put in at a constant pressure of $\sim 0.1$ MPa. 
The solution was kept at $-40 ^\circ$C. 
The reaction can be recognized complete when the solution became colorless and the product precipitated. 
Then, we removed the liquid NH$_3$ by dynamically pumping the glass tube, and obtained CsO$_2$ powder. 
The color of CsO$_2$ powder is dark yellow. 
The X--ray powder diffraction (XRPD) patterns of the samples were measured with synchrotron radiation at BL--8A of KEK--PF (wave length $\lambda = 0.99917 \AA$). 
Rietveld refinement was performed to obtain the structural parameters using the GSAS II package \cite{GSAS}. 
The final weighted $R$--factor, $R_{\rm wp}$, for room temperature structure was converged to 4.36 \%, indicating a good fit to the experimental data. 
The magnetization, $M$, was measured using a SQUID magnetometer (MPMS--R2 and MPMS3, Quantum Design Co. Ltd.) 
in the temperature region $> 2$ K. 
High magnetic field magnetization was measured in pulsed magnetic fields up to 60 T below 1.3 K. 
Because the CsO$_2$ sample is very sensitive to the atmosphere, the samples must be handled in the Ar--filled glove box. 

\begin{figure}
\begin{center}
\includegraphics[clip, width= 0.4 \textwidth]{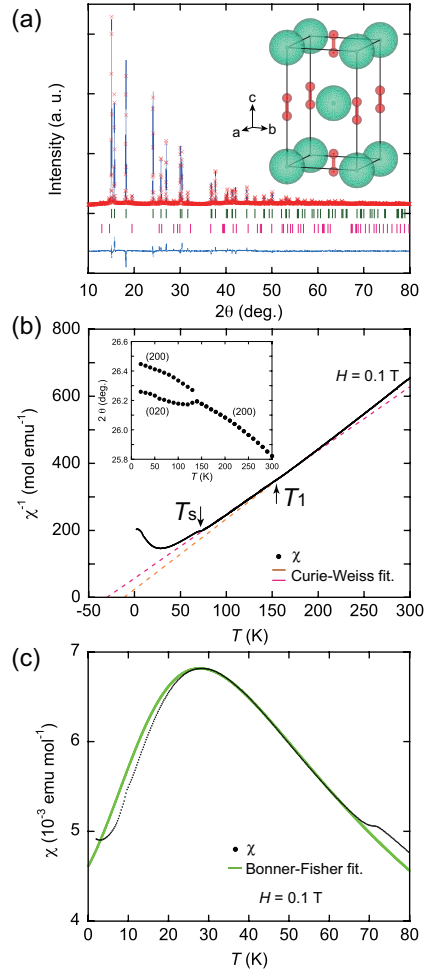}
\caption{
(a) Room temperature X--ray powder diffraction pattern collected from CsO$_2$ using $\lambda = 0.99917 \AA$. 
Observed data (red crosses), the calculated pattern (blue line), the candidate peak positions (light green bar for CsO$_2$ 
and pink bar for CsOH$\cdot$H$_2$O) and the difference between the observed and the calculated data (light blue line) are shown. 
The inset depicts the room--temperature structure of CsO$_2$, where Cs and O atoms are shown by the green and the red spheres, respectively. 
(b) Temperature dependence of inverse magnetic susceptibility, $\chi^{-1}$, of CsO$_2$ (black dots) with Curie--Weiss fits (pink and orange dotted lines). 
The fitted parameters are described in the text. 
(c) Temperature dependence of $\chi$ (black dots) below $T_{\rm S}$ with a Bonner--Fisher fit (light green line).  
The anomaly at $T_{\rm{N}} = 9.6$ K corresponds to the AF ordering. 
}
\label{XRDandM}
\end{center}
\end{figure}

%\section{Results and Discussion}
Figure \ref{XRDandM} (a) shows the XRPD pattern of CsO$_2$ at room temperature. 
All peaks can be indexed by the tetragonal symmetry (I4/mmm). 
Very small amount of impurity was confirmed to be CsOH$\cdot$H$_2$O, which fraction was estimated to be less than 3 \% 
from the XRPD analysis. 
This impurity phase should not influence the intrinsic magnetic property of CsO$_2$ because it has no unpaired electrons. 
The inset of figure \ref{XRDandM} (a) depicts the schematic figure of the crystal structure for the room temperature phase of CsO$_2$.  
The lattice parameter of CsO$_2$ is estimated to be $a = 4.469 \AA$ and $c = 7.324 \AA$, which is consistent with the literature \cite{Blake}. 
The O--O distance is estimated to be $1.11 \AA$, which may be effectively reduced by the libration of O$_2$ molecule. 
%%%%%%
%%%%%%

Figure \ref{XRDandM} (b) shows the temperature dependence of inverse magnetic susceptibility, $\chi ^{-1}$, of CsO$_2$ under a field cooling condition. 
The applied magnetic field is 0.1 T. 
No Curie--tail at low temperatures indicates high--quality of this sample. 
In the paramagnetic region, 
from room temperature to $T_1 \sim 150$ K, the $\chi$ follows Curie--Weiss law with Weiss temperature, $\theta = -10.1$ K, 
and effective magnetic moment, $\mu_{\rm eff} =1.95 \mu_{\rm B}$ while from $T_1$ to $T_{\rm S} \sim 70$ K,  $\theta = -30.0$ K and $\mu_{\rm eff} = 2.05 \mu_{\rm B}$, 
where $\mu_{\rm B}$ is Bohr magneton. 
$T_{\rm S}$ corresponds to the reported structural phase transition temperature \cite{Blake}. 
The $\mu_{\rm eff}$ is slightly larger than the value expected from spin--$\frac{1}{2}$, suggesting the orbital contribution. 
These values of $\theta$ and $\mu_{\rm eff}$ are consistent with the previous results \cite{Blake, Zumsteg}.
We note that the change in $\chi$ around $T_1$ can be recognized in a different sample batch. 
Accordingly, the AF interaction among the spins on O$_2$ molecules becomes more dominant below $T_1$. 
In the inset of figure \ref{XRDandM} (b), 
the (200) Bragg reflection in the high--temperature tetragonal phase splits into two peaks with (200) and (020) indices below 150 K, 
suggesting a tetragonal--to--orthorhombic structural change around 150 K. 
Thus, the structural phase transition at $T_1$ should be related with the enhancement of AF interactions. 
We will discuss the structural study in a separated paper in detail. 

Below $T_{\rm S}$, the $\chi$ shows a broad maximum around 28 K, indicating a low--dimensional character of the spin system. 
As shown in the figure \ref{XRDandM} (c), the $\chi$ below $T_{\rm S}$ can be fitted by the well--known Bonner--Fisher formula \cite{BF}. 
When we write the antiferromangetic Heisenberg Hamiltonian as 
$\displaystyle {\mathcal{H} = -J_{\rm{1D}} \sum_i \bm{S_i}\cdot \bm{S_{i+1}} } $, 
the 1D AF interaction, $J_{\rm{1D}}/k_{\rm{B}}$, is estimated to be  $42.8$ K, 
where $k_{\rm{B}}$ is a Boltzmann constant. 
The obtained $J_{\rm{1D}}/k_{\rm{B}}$ is consistent with the previous results \cite{Blake, Klanjsek}. 
We could not observe the shift of temperature, at which the $\chi$ showed the maximum,  under the magnetic field up to 7 T. 
The shift is evidenced in the TLL phase in the  spin--$\frac{1}{2}$ AF Cu(C$_4$H$_4$N$_2$)(NO$_3$)$_2$ \cite{kono}. 
%%%%%%
\begin{figure}
\begin{center}
\includegraphics[width=0.4 \textwidth]{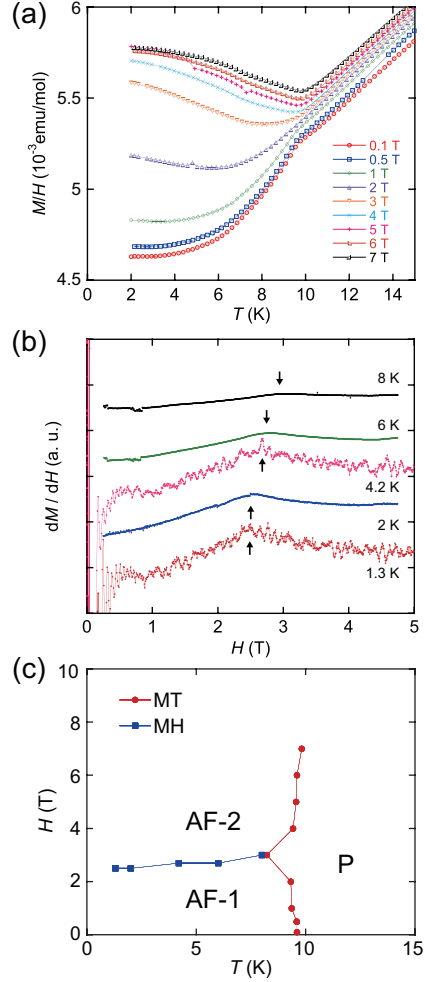}
\caption{
(a) Temperature dependence of $M/H$ for CsO$_2$ at several magnetic fields. 
(b) Isothermal derivative magnetization data, d$M$/d$H$, as a function of magnetic field below $T_{\rm{N}}$. 
The data at 1.3 K and 4.2 K were obtained by the pulsed magnetic field experiments while the others were obtained by differentiating the MPMS data. 
The arrows point at anomaly in the d$M$/d$H$ plot for different temperatures, indicating the spin--flop transition fields. 
The data are shifted for clarity. 
(c) A possible phase diagram of CsO$_2$, where P, AF--1 and AF--2 represent a paramagnetic, an antiferromagnetic--1 and an antiferromagnetic--2 phase, respectively. 
The red circles are obtained from the temperature dependence of magnetization (MT) while 
the blue squares are obtained from the magnetic field dependence of magnetization (MH). 
}
\label{SF}
\end{center}
\end{figure}
%%%%%

The anomaly at $T_{\rm{N}} = 9.6$ K corresponds to the AF ordering \cite{Blake}. 
Figure \ref{SF} (a) shows the temperature dependence of $M/H$ at several magnetic fields. 
When the magnetic field was increased above 3 T, the $M/H$ increased at low temperatures, suggesting the existence of a spin--flop transition around 3 T. 
Figure \ref{SF} (b) shows the isothermal derivative magnetization data, d$M$/d$H$, as a function of magnetic field around $T_{\rm N}$. 
The anomalies around 3 T (arrows in the figure \ref{SF} (b) ) are clearly observed and shifted higher fields side with increasing temperatures. 
In pulsed magnetization experiments, the magnetization curves also indicate the magnetic phase transition around 2.49 T. 
We summarize a possible phase diagram of CsO$_2$ under the magnetic field in the figure \ref{SF}(c). 
This magnetic phase diagram is typical of the antiferromagnet with an easy--axis anisotropy, and the phase transition from the AF--1 to the AF--2 phase 
should correspond to the spin--flop transition. 
We note that no hysteresis in the magnetization curve was observed. 

\begin{figure}
\begin{center}
\includegraphics[width=0.4 \textwidth]{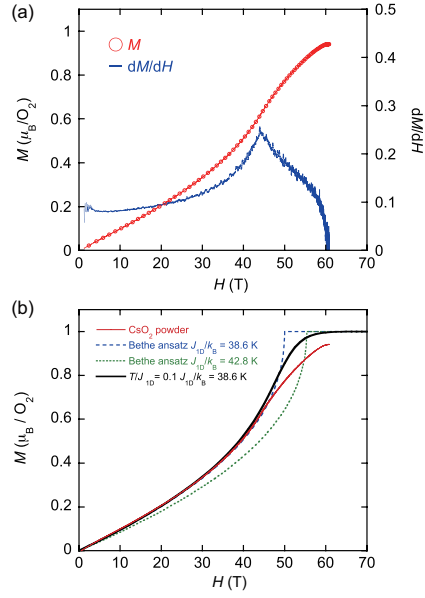}
\caption{
(a) High magnetic field magnetization, $M$, curve (red circles) and its derivative, d$M$/d$H$, curve (blue line) for CsO$_{2}$ at 1.3 K. 
In the vertical axis, $M=1$ corresponds to $1 \mu_{\rm B}$ per the O$_2$ molecule, indicating the spin--$\frac{1}{2}$.
(b) 
The dashed lines are Bethe--anzatz results for $T = 0$ with $J_{\rm 1D}/k_{\rm B} =38.6$ K (blue dashed line) and  $J_{\rm 1D}/k_{\rm B}=42.8$ 
K (green dashed line). 
The black line shows the theoretical calculation with $T/J_{\rm 1D} = 0.1$ and $J_{\rm 1D}/k_{\rm B}=38.6$ K. 
}
\label{HM}
\end{center}
\end{figure}

Figure \ref{HM} (a) shows the magnetization and its derivative curves for CsO$_2$ as a function of magnetic field up to 
60 T at 1.3 K. 
Below $T_{\rm{N}}$, remarkable up--turn curvature in the magnetization around a saturation field of $\sim 60$ T is found, 
suggesting the low--dimensional nature of this spin system. 
The saturated magnetization is also estimated to be $\sim 1 \mu_{\rm B}$, which corresponds to the spin--$\frac{1}{2}$. 
This is consistent with the magnetic susceptibility experiments. 
As shown in figure \ref{HM} (b), 
the fit with the Bethe--ansatz curve \cite{Griffiths} gives the saturation magnetization of $H_{\rm S} = 50$ T and $J_{\rm 1D}/k_{\rm B } = 38.6 $ K. 
From this, low--field magnetization can be reproduced by the exact curve, 
but, at high--field region, especially around $H_{\rm S}$, the high--field magnetization seems to be inconsistent with the calculation. 
On the other hand, if we used the $J_{\rm 1D}/k_{\rm B} = 42.8 $ K estimated from the Bonner--Fisher fit in fig. \ref{XRDandM} (c), 
the calculated magnetization did not reproduce the experiments as a whole.  
We may introduce a thermal effect and/or a higher--dimensionality of the spin system, 
i.e., an interchain coupling, to settle these inconsistency. 
The magnetization curve for temperature $T/J_{\rm 1D} = 0.1$ with $J_{\rm 1D}/k_{\rm B} = 38.6$ K in fig. \ref{HM} (b), which was calculated by a finite temperature 
DMRG \cite{okunishi}, implies better agreement around the saturation field, but could not reproduce the experiments entirely. 
Thus, the inconsistency around the saturation field in CsO$_2$ may be caused by the interchain couplings. 

%%%%%%
%%%%%%
Using the molecular field approximation, we estimate the magnetic exchange interaction and the anisotropy in CsO$_2$. 
The spin flop field, $H_{\rm SF}$, can be written by the exchange field, $H_{\rm E}$, and the anisotropy field, $H_{\rm A}$, as follows; 
%%%
\begin{equation}
H_{\rm SF} = \sqrt{2 H_{\rm E} H_{\rm A}}
\end{equation}
%%%
The $H_{\rm E}$ can be written as  $H_{\rm E} = z J S / \mathit{g} \mu_{\rm B}$, where $z$ is a number of nearest neighbor spins. 
If we assume $z = 2$ for the orbital ordered structure of CsO$_2$, which should corresponds to a 1D spin chain, 
the magnetic exchange interaction $J_{\rm 1D}/k_{\rm B} = 38.6 $ K gives the $H_{\rm E}$ of 250  kOe. 
From the figure \ref{SF}, the $H_{\rm SF}$ is estimated to be 2.49 Tesla at 2 K. 
Using these values, $H_{\rm A}$ is obtained to be 1.24 kOe. 
The $H_{\rm A}$ is almost comparable to the dipolar anisotropy field calculated in $\alpha$--phase of solid O$_2$ \cite{Uyeda}. 

Knaflic {\it et al.} found a disappearance of electron paramagnetic resonance (EPR) signals in the vicinity of $T_{\rm N}$ and 
observed an antiferromagnetic resonance (AFMR) around zero--magnetic field 
below $T_{\rm N}$ using X--band frequency \cite{Knaflic}. 
We also observed the disappearance of EPR signals in the vicinity of $T_{\rm N}$, implying a bulk long--range magnetic ordering. 
The muon spin relaxation experiments ($\mu$SR) also show clear muon--spin precession even in the zero--field condition below $T_{\rm N}$, \cite{muSR} 
proving that the ground state of CsO$_2$ under the magnetic field 
is not a field--induced ordered state, which claimed by the authors \cite{Knaflic}. 
Although we tried to measure the AFMR using the X--band frequency,  no corresponding AFMR signal were found down to 5 K. 
The AFMR relation strongly depends on the anisotropy as well as the magnetic field. 
We speculate that our measured frequency is too high compared with the zero--field excitation energy to observe the AFMR signal.
On the contrary, the observed AFMR field may not be simply explained by a general AFMR relation with an uniaxial anisotropy. 
As the EPR field at X--band frequency is about 0.3 T, which is one--order of magnitude smaller than the $H_{\rm SF}$, 
the AFMR signal at the X--band frequency may be observed at a magnetic field 
around the $H_{\rm SF}$, which is higher than the observed field, if we assume only the uniaxial anisotropy. 
Therefore, other anisotropy may be taken into consideration to explain the observation of AFMR around zero magnetic field at the X--band frequency. 
In fact, in the $\alpha$--phase of solid O$_2$, orthorhombic magnetic anisotropy was estimated from the AFMR, where 
it is suggested that the one is originated from the O$_2$ molecule itself and the other from the dipolar interaction \cite{Uyeda}. 

The magnetic exchange interactions have been calculated in the orbital--ordered KO$_2$, 
in which both the crystal field from the cations and the Coulomb interactions are thought to be dominant \cite{Kim, Solovyev}. 
Kim {\it et al.} suggested that the coherent ferro--orbital ordering of O$_2$ molecules was important to realize the experimentally observed AF structure \cite{Smith}. 
In CsO$_2$, it is proposed from the XRPD and the Raman scattering experiments that the low temperature structure 
has a $a \times 2b \times 2c$ periodicity, which may be accompanied with the coherent tilting of O$_2$ molecular axis \cite{Blake}. 
This would leads to ferro--orbital ordering of O$_2$ molecules along the [100] direction and  antiferro--orbital ordering 
along the [010] and [111] directions. 
If we consider the magnetic super--exchange interaction via Cs$^+$ ions between the nearest neighbor (NN) O$_2$ molecules 
with the framework of the Kanamori--Goodenough rule \cite{KG}, 
it may be expected that the ferro--magnetic interaction is dominant along the [100] direction 
while the antiferromagnetic interaction along the [010] and [111] directions. 
The [010] direction had been proposed as the 1D axis. 
If we refer the calculated exchange interaction in KO$_2$, 
the interchain magnetic exchange interactions between NN O$_2$ molecules may be enough to induce the long--range magnetic ordering in CsO$_2$. 
On the contrary, it has already mentioned that CsO$_2$ shows another structural phase change around $T_1 = 150 $ K. 
Thus, to understand the magnetic exchange interaction between O$_2$ molecules, the exact orbital--ordered structure at low temperature should be indispensable. 

%\section{Conclusion}
In conclusion, we have synthesized high--quality CsO$_2$ and investigated the magnetic properties. 
The obtained magnetic phase diagram is similar to that for the antiferromagnet with an easy axis anisotropy. 
High magnetic field magnetization of CsO$_2$ was performed up to 60 Tesla for the first time and exhibited remarkable up--turn curvature around  the saturation field, 
implying that CsO$_2$ is a candidate for quasi--1D spin--$\frac{1}{2}$ antiferromagnet. 
On the other hand, the theoretical calculation with $T=0$ and $T/J_{\rm{1D}}=0.1$ could not  wholly reproduce the experiment. 
To settle these inconsistency, further experiments including the low--temperature structure should be highly desirable.

%\begin{acknowledgment}
We would like to thank Prof. T. C. Kobayashi for valuable discussions on magnetism of CsO$_2$. 
The X--ray diffraction patterns were measured in research projects (2017G636) of KEK--PF. 
This research is partly supported by KAKENHI grants from Japan Society for the Promotion of Science (15H03529). 
%\end{acknowledgment}

\end{document}